\documentstyle[budapest]{article}  
\frompage{000} \topage{000}                                              
\def\rightmark{Partonic Effects} 
\def\leftmark{C. M. Ko et al.}

\title{Partonic Effects in Heavy Ion Collisions at RHIC}
\authors{
{C. M. Ko, Zi-wei Lin, and Subrata Pal%
}\\[2.812mm]
{\normalsize
\hspace*{-4pt} Cyclotron Institute and Physics Department,
Texas A\&M University,\\
College Station, Texas 77843-3366
}}
 
\abstract{Effects of partonic interactions in heavy ion collisions
at RHIC are studied in a multiphase transport model (AMPT)
that includes both initial partonic and final hadronic interactions.
It is found that a large parton scattering cross section is needed to
understand the measured elliptic flow of pions and two-pion 
correlation function.}

 
\begin{document}
 
\maketitle
\setcounter{page}{1}

\section{Introduction}

In heavy ion collisions at RHIC, a quark-gluon plasma is expected to 
be created in the initial stage. Many observables have been suggested 
as possible signals for the quark-gluon plasma. These include enhanced 
production of dileptons \cite{shuryak} and strange hadrons \cite{rafelski}, 
suppression of $J/\psi$ production \cite{satz}, Hanbury-Brown-Twiss
interferometry between pions \cite{pratt}, elliptic flow of various 
particles \cite{ollitrault}, quench of jet production \cite{wang}, 
and fluctuations in the net charge \cite{asakawa}, net baryon \cite{jeon}
as well as total baryon numbers \cite{lin}.  However, to establish these 
observables as unambiguous signals for the quark-gluon plasma requires 
that they are not affected appreciably by the final hadronic dynamics 
in the collisions. To study quantitatively these signals, we have developed 
a multiphase transport model \cite{ampt} that includes both the dynamics 
of initial partonic and final hadronic matters. With this model, we 
have studied not only the rapidity and transverse momentum distributions of
various charged particles \cite{charge} but also their elliptic flows 
\cite{flow}. Furthermore, we have investigated multistrange baryon 
production \cite{strange}, $J/\psi$ suppression \cite{jpsi} and
pion interferometry \cite{hbt} in these collisions. We find that the model 
describes very well the experimental data on the charged particle 
rapidity distributions and transverse momentum spectra. However, these 
results are more sensitive to the hadronic interactions than the partonic 
interactions. On the other hand, both the charged particle elliptic flow 
and the two-pion correlation function are affected by partonic interactions, 
and to explain the observed large elliptic flow and the measured pion 
correlation function in the experiments requires a large parton-parton 
scattering cross section in the initial partonic matter. 

\section{Multiphase Transport Model}

In the AMPT model, the initial conditions are obtained from  
the HIJING model \cite{hijing} by using a Woods-Saxon radial shape for the   
colliding nuclei and including the nuclear shadowing effect on parton
production via the gluon recombination mechanism of Mueller-Qiu 
\cite{mueller}. After the colliding nuclei pass through each other, 
the Gyulassy-Wang model \cite{gyulassy} is then used to generate the 
initial space-time information of partons. In the default HIJING model, 
these minijet partons are allowed to lose energy via the gluon 
splitting mechanism and transfer their energies to the nearby strings 
associated with initial soft interactions. Such a jet quenching 
is replaced in the AMPT model by explicitly taking into account 
parton-parton collisions via Zhang's Parton Cascade (ZPC) \cite{zpc}.
At present, only gluon elastic scatterings with a default cross
section of 3 mb are included, so the partons do not suffer any 
inelastic energy loss as they traverse the dense matter. After 
partons stop interacting, they combine with their parent strings and, 
after an average proper formation time of 0.7 fm/c, are converted to 
hadrons using the Lund string fragmentation model \cite{jetset}. 
Dynamics of the resulting hadronic matter is then described by a 
relativistic transport model (ART) \cite{art}, which has been improved 
to include baryon-antibaryon production from meson-meson interactions 
and their annihilation \cite{ko}.

The parameters in the AMPT model are determined by fitting
the experimental data from Pb+Pb collisions at center-of-mass energy 
$\sqrt s=17A$ GeV at SPS \cite{na49}.
Specifically, to describe the measured net baryon rapidity distribution,   
we have included in the Lund string fragmentation model the popcorn 
mechanism for baryon-antibaryon production with equal probabilities for 
baryon-meson-antibaryon and baryon-antibaryon configurations. 
Also, to account for both the pion yield and the enhanced kaon 
yield in this reaction, we have adjusted the two parameters in the momentum
splitting function used in the Lund string fragmentation model,
following the expectation that the string tension is modified in 
the dense matter formed in the initial stage of heavy ion collisions. 

Since the probability for minijet production is very small in heavy 
ion collisions at SPS energies, the partonic stage does not play any 
role in these collisions. We find that final-state hadronic scatterings 
reduce the proton and antiproton yields, but increase kaon and antikaon 
production by about 20\%. In contrast, the kaon yield in the default 
HIJING model is smaller than our final result by about 40\%.

\section{Rapidity and Transverse Momentum Distributions}

\begin{figure}[ht]
\insertplot{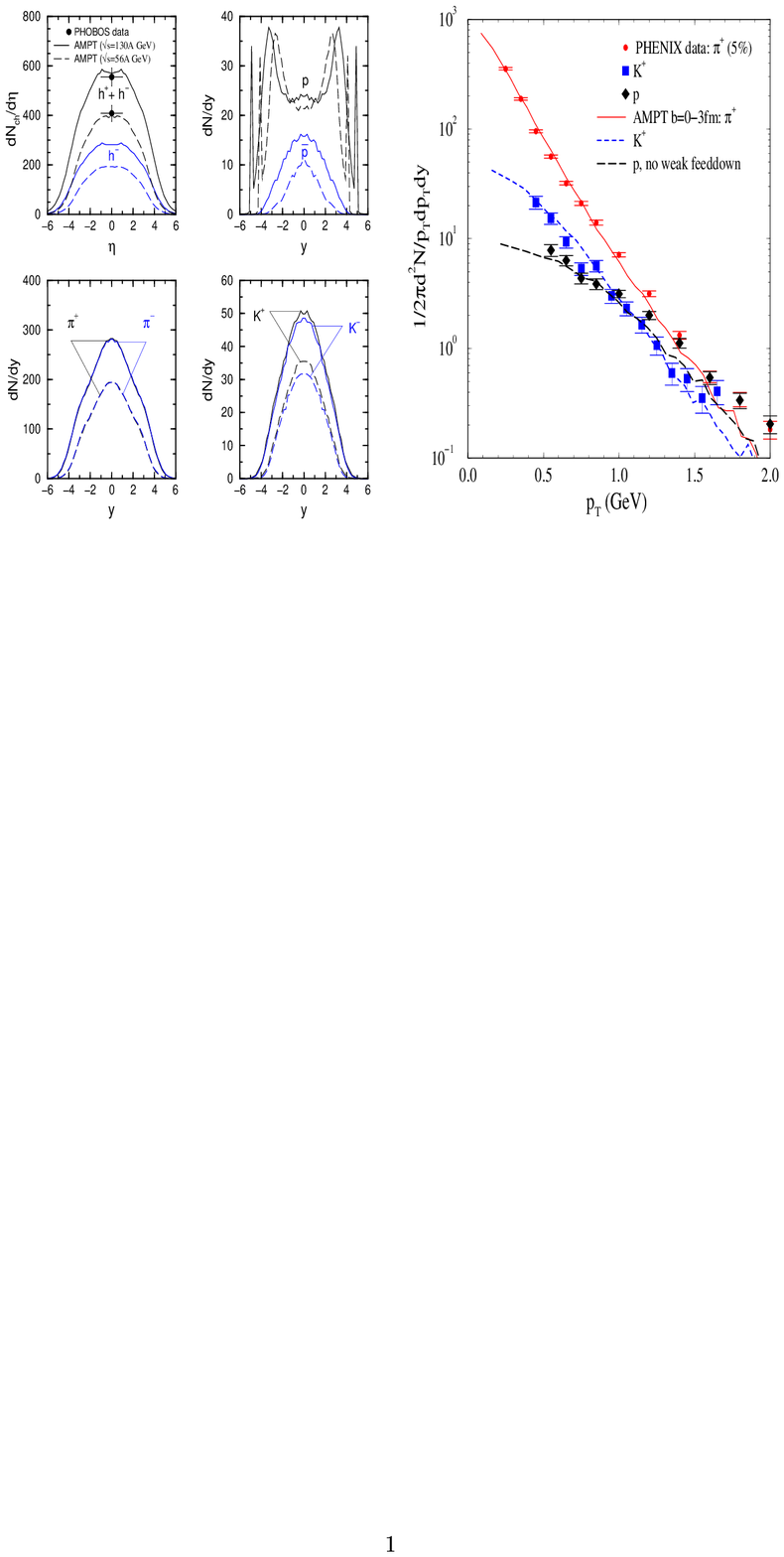}
\vspace*{-0.7cm}
\caption{Left four panels: Rapidity distributions of total and negatively 
charged particles, protons and antiprotons, charged pions, and
charged kaons in Au+Au collisions at $\sqrt s=56A$ and $130A$ GeV.  
Right panel: Transverse momentum spectra of pions, kaons, and protons
in Au+Au collisions at $\sqrt s=130A$ GeV.
The experimental data are for 5\% most central collisions from the PHOBOS 
collaboration \protect\cite{phobos} for rapidity distributions and from 
the PHENIX collaboration \protect\cite{phenix} for transverse momentum 
spectra. Results from the AMPT model shown by curves are
for impact parameters of $b\leq 3$ fm.} 
\label{rhic}
\end{figure}

Results on the charged particle rapidity distributions in central 
Au+Au collisions at RHIC are shown in the left four panels of
Fig. \ref{rhic} for center-of-mass energies of $56A$ GeV (dashed curves) 
and $130A$ GeV (solid curves) together with data from the PHOBOS 
collaboration \cite{phobos}. The measured total charged particle 
multiplicities at mid-pseudorapidity at both energies are well 
reproduced by our model. Also, the $\bar p/p$ ratio predicted by 
the AMPT model is consistent with that measured in experiments 
\cite{pbp}. The transverse momentum spectra of pions, kaons, and
protons in Au+Au collisions at $\sqrt s=130A$ GeV are shown in the 
right panel of Fig. \ref{rhic}. The results from the the AMPT model
are seen to also agree with the experimental data from the PHENIX 
collaboration \cite{phenix}.

\begin{figure}[ht]
\insertplot{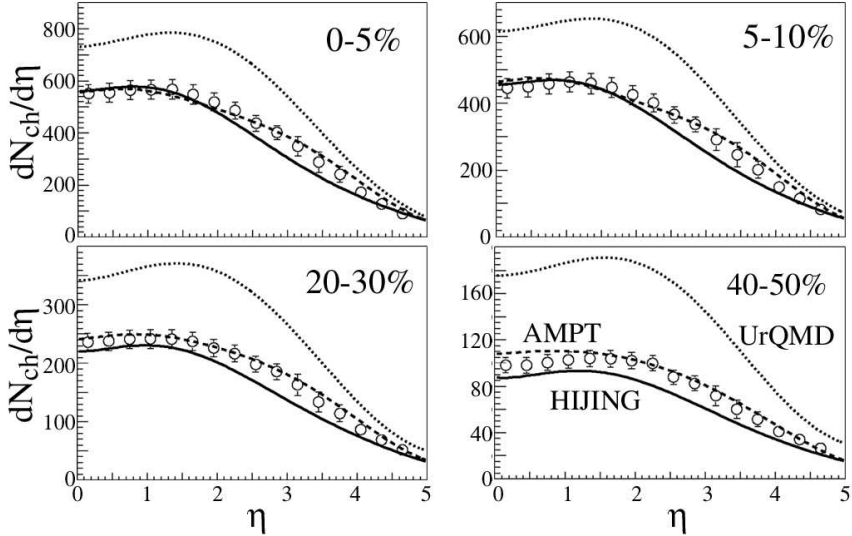}
\vspace*{-0.7cm}
\caption{Charged particle rapidity distributions 
for centralities of 0-5\%, 5-10\%, 20-30\%, and 40-50\%
in Au+Au collisions at $\sqrt{s}=130A$ GeV.
The experimental data are from the BRAHMS collaboration 
\protect\cite{brahms}. Theoretical results from the AMTP model
are shown by dashed curves while those from the default
HIJING model are given by solid curves. The figure is taken
from Ref.\cite{brahms}.} 
\label{brahms}
\end{figure}

Furthermore, the AMPT model describes very well the 
centrality dependence of the charged particle rapidity 
distribution measured by the BRAHMS collaboration \cite{brahms} as 
shown in Fig. \ref{brahms} by dashed curves. 
Compared to the default HIJING model given by solid curves, 
hadronic scatterings are responsible for the broader rapidity 
distributions seen in the experiments. 

Without hadronic interactions, there is a significant increase in the
numbers of total charged particles, pions, protons, and antiprotons
at midrapidity. The kaon number is, on the other hand, reduced slightly. 
As a result, the ratios of $\bar p/p$ and $K^+/\pi^+$ in the absence
of final-state hadronic interactions are $0.80$ and $0.13$, respectively, 
instead of $0.66$ and $0.18$ from the default AMPT model, which are
more consistent with the data.
Although the default HIJING \cite{xnwang} gives a total 
charged particle multiplicity at midrapidity that is consistent with 
the PHOBOS data, including hadronic scatterings would reduce its 
prediction appreciably.

Effects of partonic dynamics on final hadronic observables
can be studied in the AMPT model by turning off the partonic cascade.   
We find that this leads to less than $\sim 5\%$ change in the final 
charged particle yields in Au+Au collisions at $\sqrt s=130A$ GeV. 
The multiplicity distribution of hadrons is thus not very sensitive 
to parton elastic scatterings. 

\section{Elliptic Flow}

Elliptic flow in heavy ion collisions measures the anisotropy of particle
momentum distributions in the plane perpendicular to the beam direction. 
It results from the initial spatial anisotropy in non-central collisions
and is thus sensitive to the properties of the hot dense matter formed 
during the initial stage of heavy ion collisions \cite{ollitrault,rqmd}.
In transport models based only on the parton cascade, 
the elliptic flow has been shown to be sensitive to the parton scattering 
cross section, and a large value can be obtained with a large cross section 
\cite{zhang}.

\begin{figure}[ht]
\insertplot{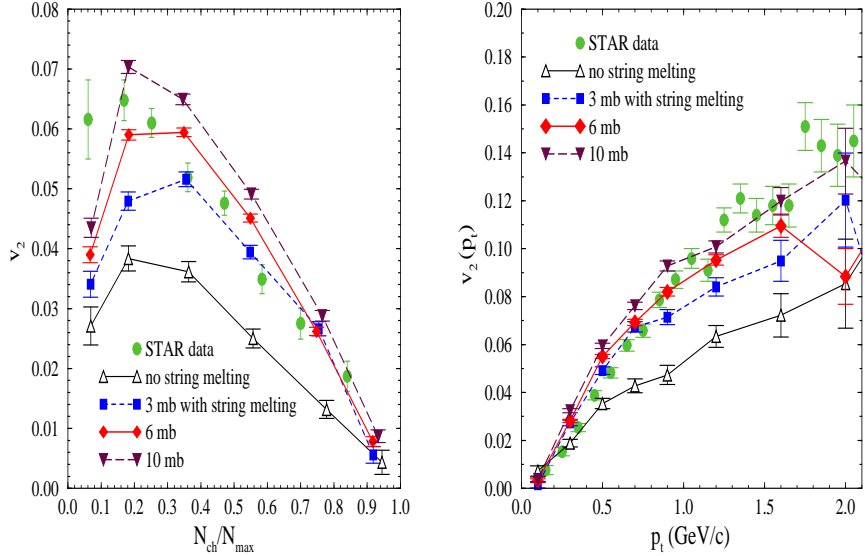}
\vspace*{-0.7cm}
\caption{Charged particle multiplicity (left panel) and transverse 
momentum (right panel) dependence of elliptic flow in Au+Au collisions
at $\sqrt{s}=130A$ GeV. Data from the STAR collaboration \protect\cite{star} 
are shown by filled circles, while theoretical results for different 
partonic dynamics are given by curves.}
\label{v2}
\end{figure}

Results on the charged particle elliptic flow from the AMPT model
for Au+Au collisions at center-of-mass energy of $130A$ GeV
are shown in Fig. \ref{v2} as a function of both the charged particle
multiplicity (left panel) and transverse momentum (right panel). 
Theoretical values from the default AMPT model (no string melting) 
are much smaller than that measured in the experiments by the STAR
collaboration \cite{star}. This is mainly due to the small fraction 
of energy that is carried by minijet partons and the lack of transverse 
collective motion of the strings. As a result, the elliptic flow is 
significantly reduced after minijet partons combine with strings and 
fragment to hadrons.

Since the initial energy density of produced matter in central 
Au+Au collisions at RHIC is more than an order of magnitude higher 
than the critical energy density ($\sim 1$ GeV/fm$^3$) expected for
the formation of the quark-gluon plasma, keeping strings in the 
high energy density region thus underestimates the partonic effects 
in these collisions.  To model the melting of strings into partons 
in regions of high energy density, we have extended the AMPT model 
to allow the initially excited strings to fragment into hadrons 
using the LUND fragmentation model and then convert these hadrons
to their constituent quarks but with bare masses, i.e., a meson is 
converted to a quark and an anti-quark, while a baryon is converted 
to three quarks.  We further assume that quarks are produced 
isotropically in the rest frame of a hadron and start to interact only 
after a proper formation time given by the inverse of the hadron transverse 
mass. Scatterings among these quarks are then treated using the parton 
cascade ZPC. After the quarks stop interacting, we model the hadronization 
by combining the nearest quark and antiquark into a meson and three quarks
into a baryon with the same flavor. The resulting hadrons are given an 
additional formation time of 0.7 fm$/c$ in their rest frame and then 
imported to the ART hadronic transport model to take into account their 
rescatterings.

With strings converted to partons, the initial energy originally 
stored in strings also contributes to the parton dynamics. 
This leads to a larger elliptic flow and also a stronger dependence
on the parton scattering cross section, making it possible to 
determine the strength of partonic interactions from the final elliptic flow.
As shown in Fig. \ref{v2}, a large parton scattering cross section
of 6-10 mb is needed to account for the observed dependence 
of charged particle elliptic flow on the total charged particle 
multiplicity and their transverse momentum.

\section{Pion Interferometry}

Particle interferometry based on the Hanbury-Brown Twiss (HBT) effect
can provide information not only on the spatial extent of an emission 
source but also on its expansion velocity and emission duration 
\cite{pratt,bertsch,rischke}. In particular, the long emission time 
as a result of the quark-gluon plasma to hadronic matter phase 
transition in the initial stage of heavy ion collisions is expected 
to lead to a much larger radius parameter along the direction 
of the total momentum of detected two particles than that perpendicular 
to both this direction and the beam direction. 

\begin{figure}[ht]
\insertplot{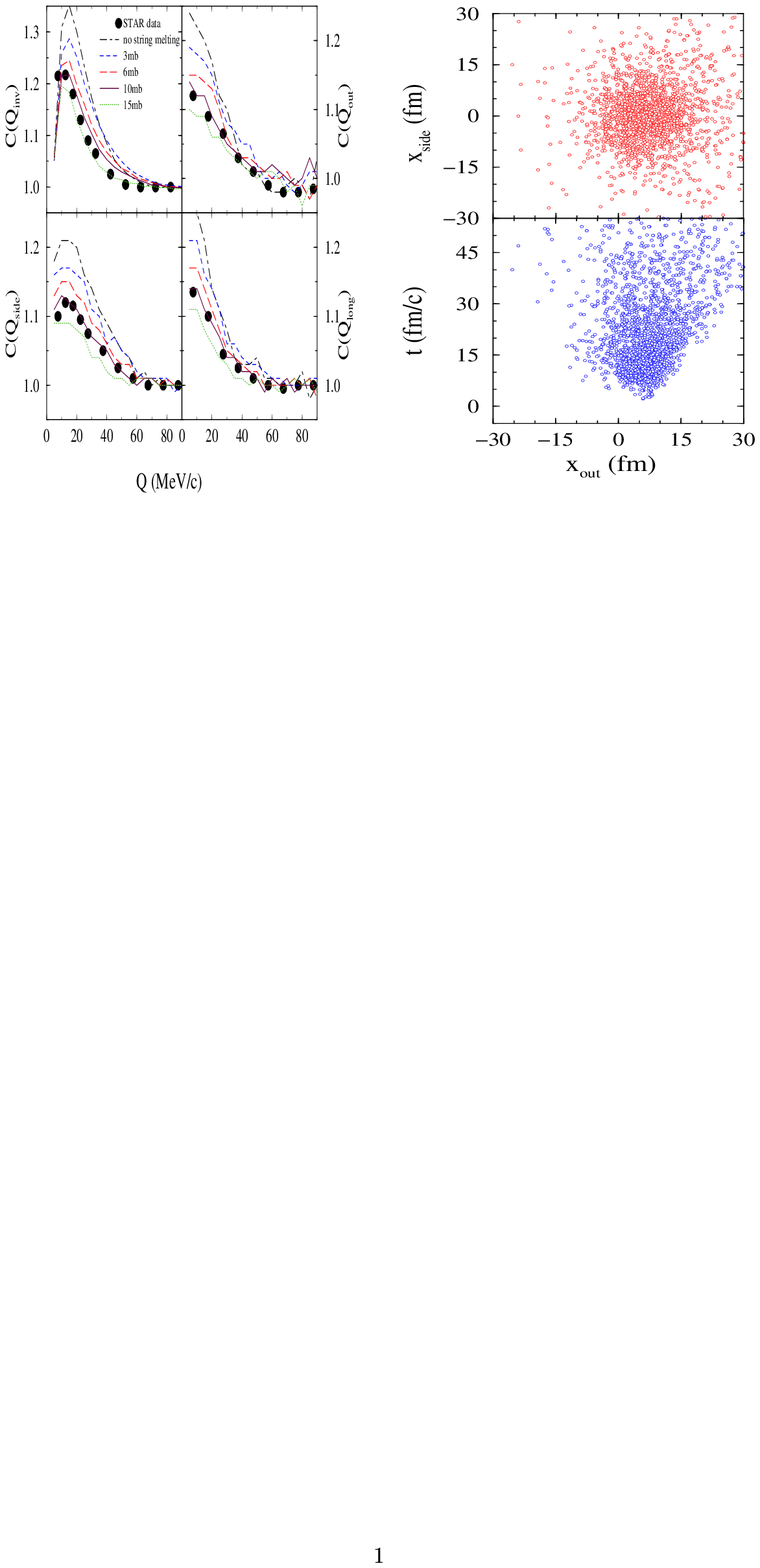}
\vspace*{-0.7cm}
\caption{Left panel:Correlation functions for midrapidity pions 
from central Au+Au collisions at $\sqrt{s}=130A$ GeV.
Coulomb-uncorrected correlation functions from the STAR collaboration 
for low $p_T$ $\pi^-$ at midrapidity from central collisions 
are shown by filled circles. Dash-dotted curves are results from the 
default AMPT model, while other curves are from the extended AMPT model 
with string melting and various values for the parton scattering cross 
section. Right panel: Distribution of the emission source 
from the AMPT model for midrapidity pions with $125<p_\perp<225$ MeV$/c$ 
in the $x_{\rm out}-x_{\rm side}$ (upper panel) and $x_{\rm out}-t$ 
(lower panel) spaces.}
\label{hbt}
\end{figure}

From the emission function, which is given by the space-time and momentum 
distribution of particles at freezeout in the AMPT model, we can evaluate 
the correlation function $C({\bf Q},{\bf K})$ of two identical pions 
including their Coulomb interaction. Here, ${\bf Q}$ and ${\bf K}$ are 
the relative and total momenta of the two pions. The six-dimension 
correlation function is usually shown as a function of the invariant 
momentum ($Q_{\rm inv}=\sqrt{-Q^2}$) or as a function of the projection of the 
relative momentum ${\bf Q}$ in the ``out-side-long'' (${\em osl}$) 
system \cite{pratt,bertsch}, defined by the beam direction 
($Q_{\rm long}$), the direction along the total transverse momentum of 
the two pions ($Q_{\rm out}$), and the direction 
orthogonal to the above two directions ($Q_{\rm side}$). 

In the left panel of Fig. \ref{hbt}, we show the one-dimensional 
projections of the correlation function for midrapidity pions with 
transverse momentum $125<p_\perp<225$ MeV$/c$ in Au+Au collisions 
at $\sqrt{s}=130A$ GeV. Also shown is the measured $\pi^-$ correlation 
function by the STAR collaboration without correcting the effect due 
to Coulomb interaction \cite{ackermann}. In evaluating the one-dimensional 
projection of the correlation function onto one of the $Q_{\rm out}$, 
$Q_{\rm side}$, $Q_{\rm long}$ axes, we have integrated the other two 
${\bf Q}$ components over the range $0-35$ MeV/$c$.  Dash-dotted curves 
in the figure are from the default AMPT model with a parton scattering 
cross section of $\sigma_{\rm p}=3$ mb, while other curves are from the 
extended AMPT model with string melting but different values of parton 
scattering cross section. It is seen that with string melting both the 
width of the $Q_{\rm inv}$ correlation function and its height decrease 
with increasing $\sigma_{\rm p}$. To reproduce the measured one-dimensional 
correlation functions from the STAR collaboration, we need a parton 
scattering cross section of about 10 mb in the extended AMPT 
model with string melting.

In the right panel of Fig. \ref{hbt}, we show the distribution of
the emission source from the AMPT model for midrapidity pions with 
$125<p_\perp<225$ MeV$/c$ in the $x_{\rm out}-x_{\rm side}$ 
and $x_{\rm out}-t$ spaces. It is seen that the emission source shows 
a large halo around a central core. The halo consists of not only pions 
from decays of long-lived resonances such as the $\omega$ but also 
thermal pions. The latter contribution to the halo becomes increasingly 
important when the collective expansion velocity of the source is large 
as a result of increasing parton cross section. The emission source also
shows a strong correlation between $x_{\rm out}$ and $t$, with 
the width of its $x_{\rm out}$ distribution increasing with 
the emission time $t$.

\begin{figure}[ht]
\insertplot{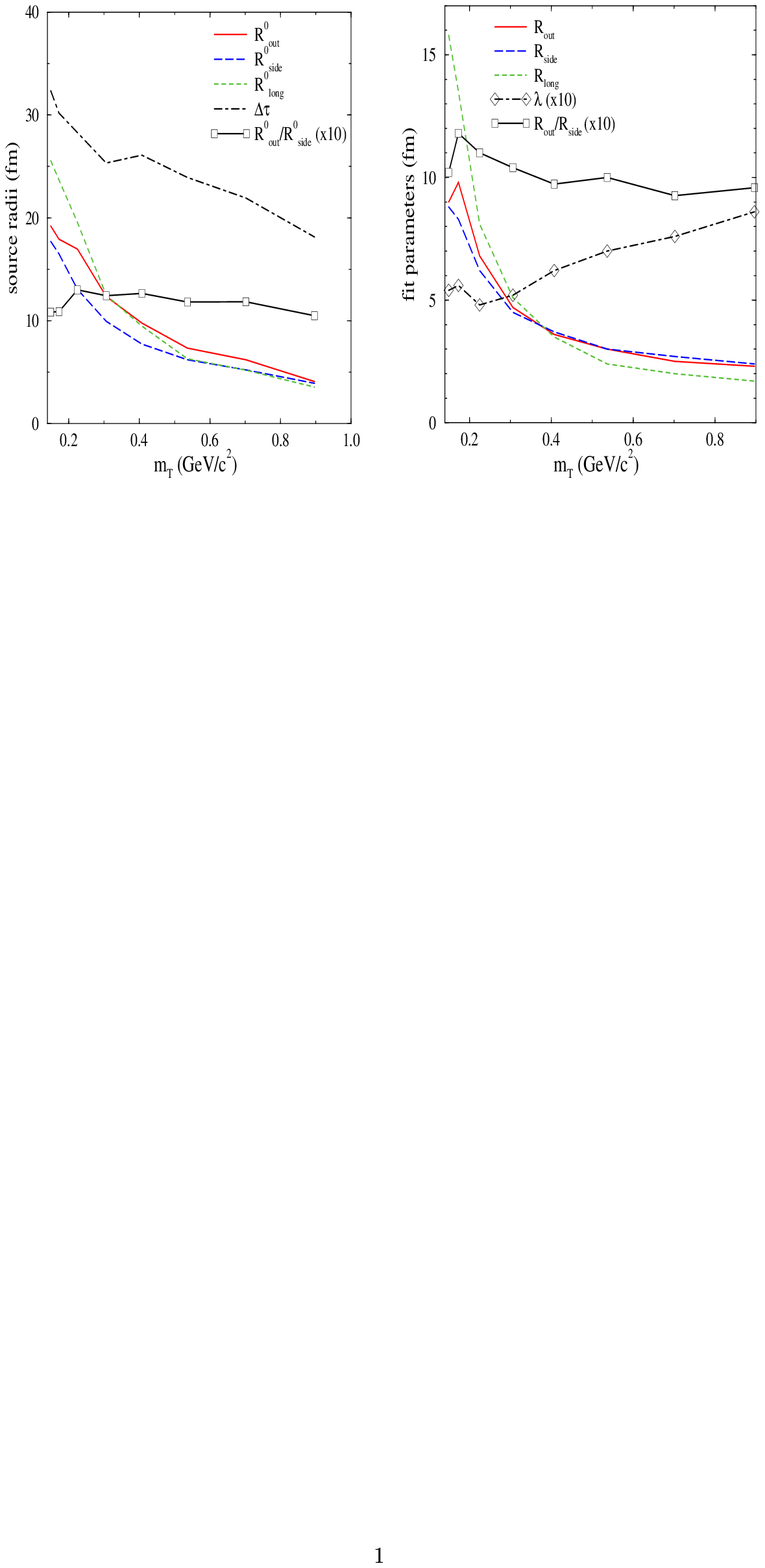}
\vspace*{-0.7cm}
\caption{Left panel: Radii $R^0_{\rm out}$, $R^0_{\rm side}$, 
and $R^0_{\rm long}$ and emission duration $\Delta\tau$ of the emission 
source for  midrapidity pions with $125<p_\perp<225$ MeV$/c$ 
in central Au+Au collisions at $\sqrt{s}=130A$ GeV as functions of 
the average transverse mass of the pion pair from the AMPT model with 
string melting and parton scattering cross section $\sigma_{\rm p}=15$ mb. 
Right panel: Same as left panel for the radius parameters obtained
from fitting the two-pion correlation by a Gaussion function in 
their relative momentum.}
\label{radius}
\end{figure}

The size of emission source can be estimated from the curvature of 
the correlation function at ${\bf Q}=0$, i.e, the second moments of
the emission function from the AMPT model. In the left panel of Fig. 
\ref{radius}, we show the dependence of the source radii of midrapidity 
pions ($-0.5<y<0.5$) on the average transverse mass $m_{\rm T}$ of the 
pion pair from the AMPT model with string melting and a parton cross section 
of 10 mb.  At low transverse mass, all three radii $R^0_{\rm out}$, 
$R^0_{\rm side}$, and $R^0_{\rm long}$ are large and have values between 
10 and 25 fm/c. The emission duration $\Delta\tau$ also lasts a long 
time of more than 25 fm/c. As the average pion transverse mass increases, 
both the emission duration and the source radii, particularly 
$R^0_{\rm long}$, are found to decrease. The ratio 
$R^0_{\rm out}/R^0_{\rm side}$ has a value of about 1.2 for 
most values of $m_{\rm T}$ and is smaller than the prediction based on the 
hydrodynamical model with freezeout treated via the hadronic transport
model \cite{soff}. 

Empirically, radii of the emission source are obtained from 
fitting the measured correlation function by a Gaussian 
function in the relative momentum of the pion pair. Results
from such a fit to the correlation function obtained from 
the AMPT model with string melting and $\sigma_{\rm p}=10$ mb are shown 
in the right panel of Fig. \ref{radius} as functions of the average
pion transverse mass. The fitted radius parameters
$R_{\rm out}$, $R_{\rm side}$, and $R_{\rm long}$ are about a factor 
of two to three smaller than corresponding source radii evaluated 
directly from the emission function, indicating that the emission source 
from the AMPT model deviates appreciably from a Gaussian one. 
Furthermore, the ratio $R_{\rm out}/R_{\rm side}$ of the 
fitted Gaussian radius parameters is much closer to one than that
from the source radii obtained from the emission function.

\section{Conclusions}

Using a multiphase transport model, which includes 
both initial partonic and final hadronic interactions, we have studied 
the rapidity distributions of charged particles such as protons,  
antiprotons, pions, and kaons in heavy ion collisions at RHIC.
With the model parameters constrained by data from central Pb+Pb collisions 
at $\sqrt s=17A$ GeV at SPS, theoretical results 
on the total charged particle multiplicity at midrapidity in 
central Au+Au collisions at $\sqrt s=56A$ and $130A$ GeV 
agree quite well with the data from the PHOBOS collaboration.
Also, the predicted $\bar p/p$ ratio is consistent with that from
the STAR and BRAHMS collaborations. Furthermore, 
the centrality dependence of the charged particle rapidity
distribution measured by the BRAHMS collaboration is well reproduced.
These hadronic observables are, however, less sensitive to 
the initial partonic interactions than to the final hadronic interactions. 
On the other hand, partonic effects are important for both the elliptic 
flow of charged particles and the two-pion correlation function.
To reproduce the measured elliptic flow and two-pion correlation function
by the STAR collaboration, we need not only to convert 
the initial strings to partons but also to use a large parton-parton 
scattering cross section. Our results thus demonstrate the possibility
of studying the dynamics of partonic matter in heavy ion collisions 
at RHIC. To further test these partonic effects, it is necessary
to study their dependence on the initial parton distribution, 
the inelastic scatterings in the partonic stage, and the 
mechanism for melting the initial strings as well as for 
hadronization. Such an improved study based on the AMPT model 
will be very useful in  finding the signals for the 
quark-gluon plasma formed in heavy ion collisions at RHIC.

\section*{Acknowledgments}
This talk is based on work supported by the National Science Foundation 
under Grant Nos. PHY-9870038 and PHY-0098805, the Welch Foundation under 
Grant No. A-1358, and the Texas Advanced Research Program 
under Grant No. FY99-010366-0081.

\end{document}